\documentclass[12pt]{iopart}

\usepackage{amssymb}
\usepackage[caption=false]{subfig}
\usepackage{graphicx}
\usepackage{iopams}
\usepackage{tikz}
\usepackage{cleveref}
\usepackage{xcolor}
\usepackage[switch]{lineno}

\usetikzlibrary{calc, fit, automata, positioning, arrows, shapes}

\crefname{equation}{}{}
\crefname{enumi}{}{}

\begin{document}

\title{Viterbi Decoding of CRES Signals in Project 8}

\author{A.~Ashtari~Esfahani$^1$, Z.~Bogorad$^2$, S.~B\"oser$^3$, N.~Buzinsky$^2$, C.~Claessens$^1$, L.~de~Viveiros$^4$, M.~Fertl$^3$, J.~A.~Formaggio$^2$, L.~Gladstone$^5$, M.~Grando$^6$, M.~Guigue$^7$, J.~Hartse$^1$, K.~M.~Heeger$^8$, X.~Huyan$^6$, J.~Johnston$^2$, A.~M.~Jones$^6$, K.~Kazkaz$^9$, B.~H.~LaRoque$^6$, M.~Li$^2$, A.~Lindman$^3$, C.~Matth\'e$^3$, R.~Mohiuddin$^5$, B.~Monreal$^5$, J.~A.~Nikkel$^8$, E.~Novitski$^1$, N.~S.~Oblath$^6$, J.~I.~Pe\~na$^2$, W.~Pettus$^{10}$, R.~Reimann$^3$, R.~G.~H.~Robertson$^1$, G.~Rybka$^1$, L.~Salda\~na$^8$, M.~Schram$^6$, P.~L.~Slocum$^8$, J.~Stachurska$^2$, Y.-H.~Sun$^5$, P.~T.~Surukuchi$^8$, A.~B.~Telles$^8$, F.~Thomas$^3$, M.~Thomas$^6$, T.~Th\"ummler$^{11}$, L.~Tvrznikova$^9$, W.~Van~De~Pontseele$^2$, B.~A.~VanDevender$^{1,6}$, T.~E.~Weiss$^8$, T.~Wendler$^4$, E.~Zayas$^2$, A.~Ziegler$^4$} 

\vspace{5mm}

\address{$^1$Center for Experimental Nuclear Physics and Astrophysics and Department of Physics, University of Washington, Seattle, WA 98195, USA}
\address{$^2$Laboratory for Nuclear Science, Massachusetts Institute of Technology, Cambridge, MA 02139, USA}
\address{$^3$Institut f\"ur Physik, Johannes-Gutenberg Universit\"at Mainz, 55128 Mainz, Germany}
\address{$^4$Department of Physics, Pennsylvania State University, University Park, PA 16802, USA}
\address{$^5$Department of Physics, Case Western Reserve University, Cleveland, OH 44106, USA}
\address{$^6$Pacific Northwest National Laboratory, Richland, WA 99354, USA}
\address{$^7$Laboratoire de Physique Nucl\'eaire et de Hautes \'Energies, Sorbonne Universit\'e, Universit\'e de Paris, CNRS/IN2P3, Paris, France}
\address{$^8$Wright Laboratory, Department of Physics, Yale University, New Haven, CT 06520, USA}
\address{$^9$Lawrence Livermore National Laboratory, Livermore, CA 94550, USA}
\address{$^{10}$Department of Physics, Indiana University, Bloomington, IN, 47405, USA}
\address{$^{11}$Institute for Astroparticle Physics, Karlsruhe Institute of Technology, 76021 Karlsruhe, Germany}

\date{\today -- v2.0}

\vspace{10pt}

\begin{abstract}
Cyclotron Radiation Emission Spectroscopy (CRES) is a modern approach for determining charged particle energies via high-precision frequency measurements of the emitted cyclotron radiation.
For CRES experiments with gas within the fiducial volume, signal and noise dynamics can be modelled by a hidden Markov model.
We introduce a novel application of the Viterbi algorithm in order to derive informational limits on the optimal detection of cyclotron radiation signals in this class of gas-filled CRES experiments, thereby providing concrete limits from which future reconstruction algorithms, as well as detector designs, can be constrained.
The validity of the resultant decision rules is confirmed using both Monte Carlo and Project 8 data.

\end{abstract}
\vspace{2pc}


\section{Introduction}

Cyclotron Radiation Emission Spectroscopy (CRES) is a recently-demonstrated technique for high-resolution energy measurements of individual charged particles \cite{artMonrealOG,artPRL0}. When placed within a magnetic field, charged particles travel in helical trajectories, thereby emitting cyclotron radiation. This radiation can serve as a sensitive, non-destructive probe into the kinematic properties of the radiating particle.

In a uniform magnetic field $B$, a charged particle with kinetic energy $K_e$ will emit cyclotron radiation with angular frequency

\begin{equation}
        \omega_c = \frac{q B}{m \gamma} = \frac{qB}{m + K_e}
        \label{eqn:cyc}
\end{equation}

\noindent where $\gamma$ is the Lorentz factor, and $q$ and $m$ are the charge magnitude and mass of the particle, respectively. Therefore, for known magnetic fields, accurate frequency measurements of the emitted radiation map into a high-resolution measurement of the kinetic energy.

While CRES can have many potential use cases \cite{artKareem}, the discussion presented here will be in the context of Project 8 \cite{artP8IOP}, as this is the current largest application of the technique.
Project 8 will attempt to determine the absolute mass scale of the neutrino via spectroscopic measurement of electrons near the endpoint of the beta decay spectrum of gaseous tritium.
All currently published implementations of CRES experiments contain gas within the fiducial volume as a source of radiating electrons.

CRES data are time series of voltages. These voltages are induced by the electromagnetic radiation incident to receiver elements, and then amplified, downmixed, and digitized.
The data consists of both Johnson-Nyquist noise, resulting from thermal motion of charge carriers in the amplification chain, and if an electron is present, a coherent cyclotron radiation signal. CRES signals are modelled by linear frequency-modulated chirps \cite{bookRadarDetection} of the form: $s(t) = A \exp{i(\phi + \omega_0 t + \alpha t^2/2)}$, as the radiation of electron energy yields a linearly increasing instantaneous cyclotron frequency (Equation \ref{eqn:cyc}: $\alpha = \omega_c P_e / (m \gamma)$).
Assuming the power released via the electron's radiation, $P_e$, is known, application of the dechirping operator ($e^{-i \alpha t^2/2}$) to the recorded CRES signal converts it to a time-limited monofrequency sinusoid.
Data is typically represented in time-frequency space via the short-time Fourier transform.

One of the primary challenges in CRES experiments is the reliable detection of signals over thermal noise, given that the power radiated via cyclotron emission is relatively weak.
In a tritium beta decay experiment operating at a magnetic field of 1 T, only $\approx$1 fW is radiated by a given 18.6 keV endpoint electron.
While efficiently detecting such events with minimal latency is critical to maximizing the sensitivity reach of the experiment \cite{artP8Sensitivity}, one must simultaneously minimize the number of false positives, which occur when noise fluctuations are incorrectly reconstructed as genuine electron events.
In the context of Project 8, false or mis-reconstructed events constitute the dominant background, with a single false positive near the tritium endpoint energy having the potential of jeopardizing the experimental sensitivity of a multi-year discovery run.

In this document, we apply the framework of detection theory to CRES signals in gas-filled detectors.
While the signal phenomenology can have additional spectral complexity within specific detector and magnetic field designs \cite{artPhenoPaper}, the basic requirements of a CRES experiment impose several key features onto the resultant data, which lead to unique analytic decision rules and detection properties.

\section{Viterbi Algorithm \& Hidden Markov Models}\label{sec:hmm}

In this section, we discuss a novel application of the Viterbi algorithm \cite{artViterbiOG,bookLightWavePapen} for generic CRES signal detection.
The Viterbi algorithm is a well-established dynamic programming method, with applications in diverse fields such as natural language processing, bioinformatics, and telecommunications \cite{bookFSNLP,bookRNAViterbi,bookFundsConvCodes}.
Fundamentally, the algorithm allows one to reconstruct the most probable underlying state sequence given time series data.

\begin{figure}[t]
\centering
\begin{tikzpicture}[->,>=stealth',shorten >=1pt,auto,node distance=5cm, semithick]
  \tikzstyle{every state}=[thick,fill=black!10]
  \tikzstyle{box}=[rectangle, draw=black!100]
    \node[state] (h0) {$\mathcal{H}_0$};
    \node[state, right of=h0] (h1) {$\mathcal{H}_1$};
    \node[state,fill=white,text=black, below of=h0] (q0) {$0$};
    \node[state, fill=black,text=white,below of=h1] (q1) {$1$};
    \node[state, right of=h0] (h1) {$\mathcal{H}_1$};
    \draw   (h0) edge[loop above] node{$T_{00}$} (h0)
        (h1) edge[loop above] node{$T_{11}$} (h1)
        (h0) edge[bend left, above] node{$T_{01}$} (h1)
        (h1) edge[bend left, above] node{$T_{10}$} (h0)
        (h0) edge node[fill=white, anchor=center, pos=0.23] {$1-p_0$} (q0)
        (h0) edge node[fill=white, anchor=center, pos=0.27] {$p_0$} (q1)
        (h1) edge node[fill=white, anchor=center, pos=0.26] {$1-p_1$} (q0)
        (h1) edge node[fill=white, anchor=center, pos=0.25] {$p_1$} (q1);
        \draw [-,loosely dashed, shorten >=-1cm, shorten <=-3cm] ($(h0)!0.5!(q0)$) coordinate (a)  --  ($(h1)!(a)!(q1)$);
        \node at (-2.2,-1.9){Hidden};
        \node at (-2.2,-3) {Observed};
\end{tikzpicture}
 \caption{State diagram of Hidden Markov Model characteristic of CRES experiments. Hidden states (upper) produce observed data (lower) with different probabilities. }
\label{fig:hmm_sparseSpec}
\end{figure}
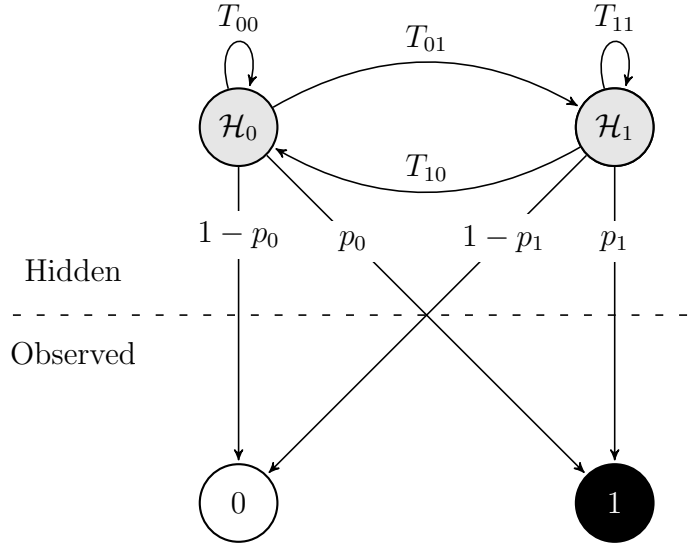

The application of the Viterbi algorithm to CRES data is motivated by re-expressing the data sequence as a so-called hidden Markov model (HMM) \cite{artHMMIntro, bookinfHMM}, as illustrated in Figure \ref{fig:hmm_sparseSpec}.
By definition, in a Markov model, the state of the system at a given time depends only on the state immediately preceding it.
In a HMM, there is an underlying stochastic Markov model controlling the sequence of states, which are hidden from the observer.
Each hidden state (upper) then emits an observable symbol (lower), drawn from a known probability distribution function.
Given the observation of a sequence of emitted symbols as a result of a HMM, the Viterbi algorithm reconstructs the most probable sequence of hidden states.
The algorithm has asymptotic time complexity $\mathcal{O}(S^2 \times D)$, where $S$ is the number of hidden states and $D$ is the number of observed data samples.

In CRES experiments, we claim that the data can generally be represented in terms of a HMM, between hidden states $\mathcal{H}_0$ and $\mathcal{H}_1$, corresponding to the noise-only and signal-plus-noise hypotheses, respectively. The transition probability $T_{ij}$ denotes the probability that the system switches from hidden state $i$ to $j$ at each discrete time sample.

\begin{figure}[t]
        \centering
            \includegraphics[width=0.85\textwidth]{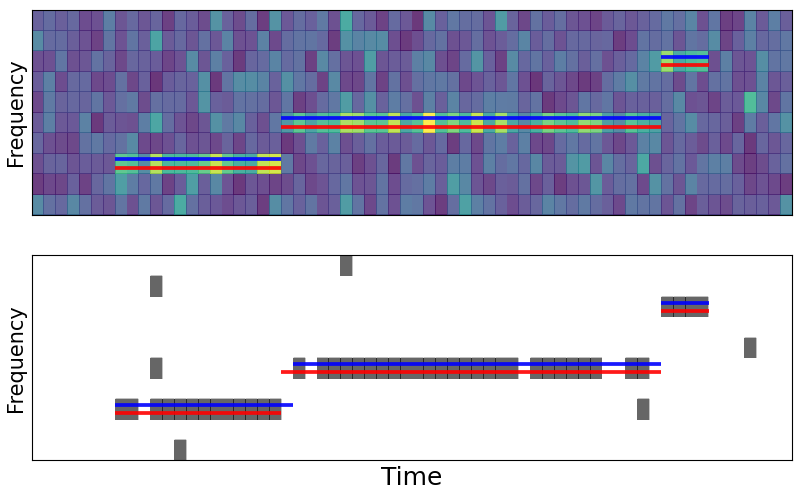}
    \caption{Example of a 3-track event generated by a HMM in short-time Fourier transform time-frequency representation, illustrated in both the raw (top) and sparse (bottom) spectrogram data formats. (Red line: MC truth; Blue line: Viterbi reconstruction). Bins have dimensions $40.96 \, \mu \textrm{s} \times 24.4$ kHz.}
    \label{fig:vittySpectrograms}
\end{figure}

As one does not have direct knowledge about whether $\mathcal{H}_0$ or $\mathcal{H}_1$ is true at a given time, one must instead use the data, in the form of Fourier voltage amplitudes. For simplicity, we initially consider binary-quantized data, referred to as the sparse spectrogram, in which the Fourier amplitudes are compared to a quantization threshold and assigned either a 0 or a 1 (Figure \ref{fig:vittySpectrograms}: lower). In Section \ref{sec:raw}, we derive the detector performance for fully continuous, or raw, data.

Application of the Viterbi algorithm requires that data be describable by a HMM such that signal and noise state transitions satisfy Markovian time-independence. CRES data satisfy this condition due to a few key characteristics of the experimental setup, which are uncommon with respect to the majority of generic signal detection problems, such as those that arise in radar or sonar applications. 

Firstly, so long as the radioactive source gas (e.g. tritium) is properly replenished, the state transition from noise-only to signal-plus-noise ($\mathcal{H}_0 \to \mathcal{H}_1$) will be constant, as decays of the gaseous source occur spontaneously and with constant rate. Secondly, in Project 8 and in CRES experiments in general, the inverse process $\mathcal{H}_1 \to \mathcal{H}_0$, corresponding to the end of an electron event, also occurs with a time-independent rate, since the time-duration of events is overwhelmingly limited by scattering between the electrons and the residual gas within the fiducial volume \cite{thesisChristine}. For electrons within a volume of gas with uniform density, traveling at nearly constant velocity, the probability of scattering with a gas molecule so as to expel the electron from the magnetic trap is also nearly independent of time.
We can therefore see how the unique case of CRES electrons embedded within a gaseous volume allows for this novel application of the Viterbi algorithm by equating electron appearances and disappearances to state transitions of a HMM.

The Viterbi algorithm is a maximal likelihood algorithm that explicitly evaluates the probability of all possible hidden state sequences, given the observed data. As a result, it is capable of explicitly determining the exact conditions such that a data segment is classified as being caused due to a genuine signal event, as opposed to being caused by consecutive noise fluctuations. We denote the probability of transitioning from hidden state $i$ to $j$ with $T_{ij}$, and the probability of emitting a high-amplitude bin in hidden state $i$ with $p_i$. For a data segment with $M$ of $N$ bins exceeding the quantization threshold, the likelihoods for the $\mathcal{H}_0$ and $\mathcal{H}_1$ hypotheses are respectively:

\begin{eqnarray}
    &p(\bi{y} \mid \mathcal{H}_0) = T_{00}^{N+1} \cdot \left[ p_0^M (1-p_0)^{N-M} \right] \label{eqn:VitCon0}\\
    &p(\bi{y} \mid \mathcal{H}_1) = \left( T_{01} \, T_{11}^{N-1} \, T_{10} \right) \cdot \left[  p_1^M (1-p_1)^{N-M} \right] \label{eqn:VitCon1}
\end{eqnarray}
\noindent where $\bi{y}$ is a vector containing the observed (noisy) data. The maximal likelihood detection criterion ${p(\bi{y} \mid \mathcal{H}_1)} > {p(\bi{y} \mid \mathcal{H}_0)}$ can therefore be rearranged to constrain  the minimum number of high-amplitude bins ($M$) needed for CRES signal detection, as a linear function of the total number of bins in the data segment ($N$):

\begin{equation}
        M > \frac{N \left[ \ln\left( \frac{T_{00}}{T_{11}} \right) + \ln \left( \frac{1-p_0}{1-p_1} \right) \right] + \ln \left( \frac{T_{00} \cdot T_{11}}{T_{01} \cdot T_{10}} \right)}{\ln \left( \frac{p_1}{p_0} \left(\frac{1-p_0}{1-p_1}\right) \right)} \label{eqn:McondVit}
\end{equation}

Transition matrix elements are derived from the properties of the gas in the trap volume. In particular, $T_{01}$, the probability of state transition $\mathcal{H}_0 \to \mathcal{H}_1$ during a given time step is $\Gamma t_b$, where $t_b$ is the time bin length and $\Gamma$ is the underlying event rate of electrons with kinetic energy within the Fourier bin. Likewise, the matrix element $T_{10}$ corresponds to the probability that an event will end in a given time bin, which depends on the electron mean free time $\tau$ as $1-e^{-t_b/ \tau}$. Other matrix elements are fixed by the normalization constraint $\sum_j T_{ij} = 1$.

The quantization threshold itself is optimized numerically. An excessively large quantization threshold would result in $p_0, p_1 \to 0$, such that the sparse spectrogram in Figure \ref{fig:vittySpectrograms} (lower) would be completely empty. Likewise, an arbitrarily small quantization threshold would yield a completely filled (black) sparse spectrogram. The optimal quantization threshold minimizes the expected number of above-threshold bins required for detection (Equation \ref{eqn:McondVit}).

\subsection{Incorporating Event Structure}

The first extension to this simplified application of the Viterbi algorithm to CRES signal reconstruction is to include information using the signal event topology to increase sensitivity to temporally short tracks. 
In particular, CRES electron tracks are usually not isolated; due to the inelastic scattering of electrons with the residual gas, a single electron can produce multiple continuous-phase \textit{tracks}, punctuated by abrupt frequency discontinuities, which together comprise a single electron \textit{event} (Figure \ref{fig:vittySpectrograms}). Eventually, a scatter will sufficiently redirect the electron velocity vector to result in the escape of the electron from the magnetic trap, ending the event. Only the start frequency of the first track in a given event is relevant for energy spectroscopy. 

A brief high-amplitude fluctuation is more likely to be a result of a genuine CRES event if there is a definitive, long-duration signal nearby in time-frequency space than if the same transient signal was found isolated within noise.
Fully incorporating this additional information from the event topology would yield additional sensitivity to shorter tracks than otherwise possible.

Our formulation of the Viterbi algorithm can be extended to incorporate this scattering information by simply adding additional signal states, labeled $\mathcal{H}_i$, corresponding to signals within the $i^\mathrm{th}$ frequency bin of the short-time Fourier transform, with indices starting from $1$, and with $\mathcal{H}_0$ still representing the noise-only hypothesis.
The transition matrix is then extended to $(N_b +1) \times (N_b +1)$, where $N_b$ is the number of frequency bins, and  contains the cross-section weighted energy loss distribution \cite{Aseev2000} of the gas composition within the detector volume.
These scattering matrix elements $T_{ij}$ ($i,j \neq 0$) correspond to the probability that a track in frequency bin $i$ will scatter to bin $j$ within a given time step, which is a product of the scattering rate and the energy loss probability distribution function, for the relevant gas particle.
Notably, the scattering matrix element $T_{ij}$ is to first-order only dependent on the difference between $i$ and $j$, corresponding to the energy loss of the scatter.
As a result, the block matrix with elements $T_{ij}$ ($i,j \neq 0$) is approximately upper triangular, since inelastic scattering in the lab frame overwhelmingly results in decreases in the kinetic energy of the incident electron ($i>j \Rightarrow T_{ij} = 0$).

Analogous to the decision criteria given in Equations \ref{eqn:VitCon0} -- \ref{eqn:McondVit}, we can explicitly derive the Viterbi detection criterion for a short track candidate in bin $i$ with $M_i$ of $N_i$ high-amplitude bins, given that there is a track in bin $j$ immediately succeeding it. Provided that the second track is sufficiently long (Equation \ref{eqn:McondVit}) so as to exclude the noise-only hypothesis, the detection criterion is found by comparing the relative probabilities of the hidden state sequences $\mathcal{H}_0 \to \mathcal{H}_i \to \mathcal{H}_j \to \mathcal{H}_0$ and $\mathcal{H}_0 \to \mathcal{H}_j \to \mathcal{H}_0$, that is, with and without the initial track candidate:

\begin{figure}[t]
        \centering
            \includegraphics[width=0.7\textwidth]{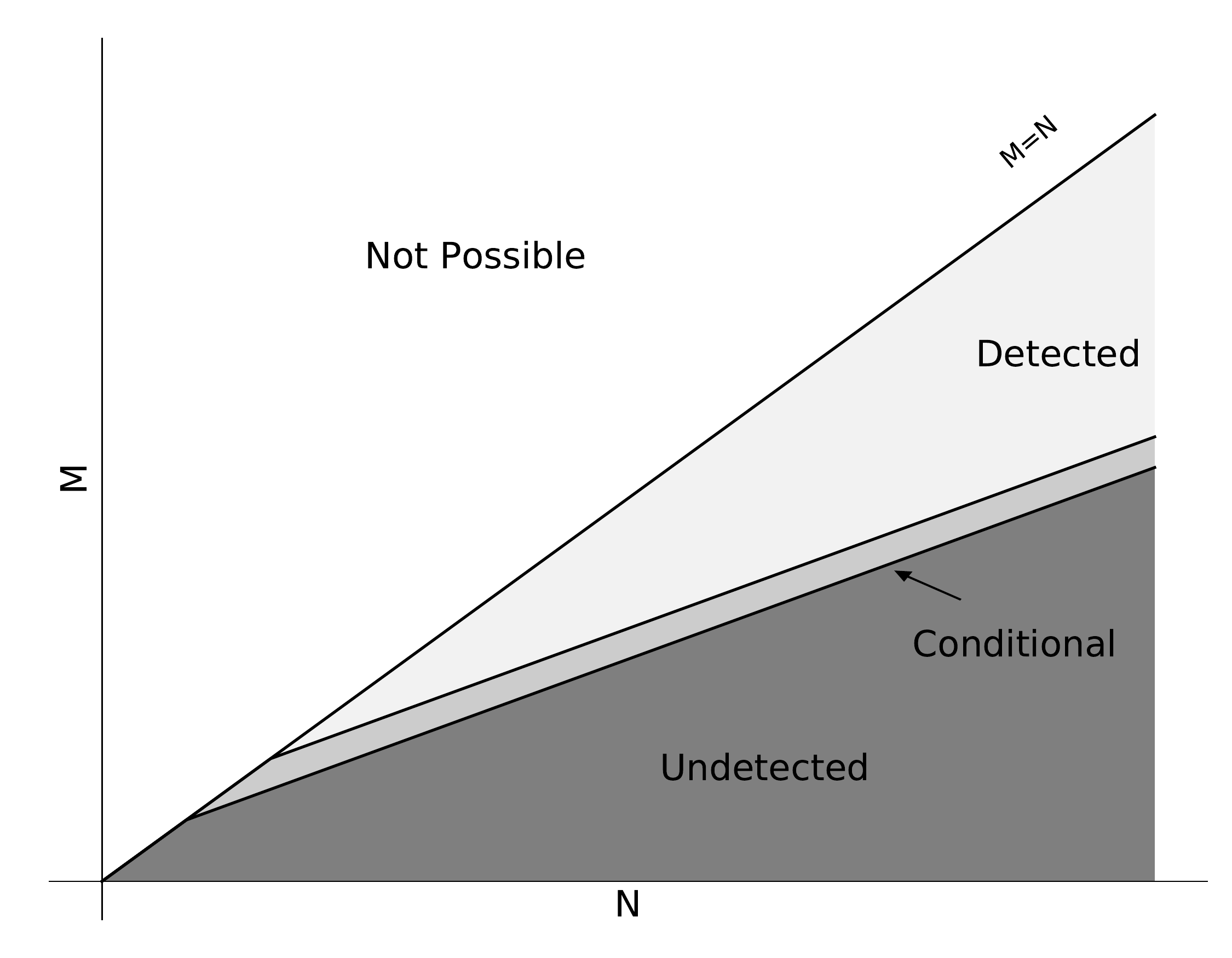}
            \caption{Illustration of detection regions for Viterbi decoding of CRES signals, as a function of the number of bins above threshold ($M$) and the total number of bins ($N$) in a track candidate. For fixed track length, the presence of a nearby event lowers the number of high-power bins $M$ required for detection, defining the conditional region (light grey).}
           \label{fig:vittyDecision}
\end{figure}

\begin{equation}
        M_i > \frac{N_i \left[ \ln\left( \frac{T_{00}}{T_{ii}} \right) + \ln \left( \frac{1-p_0}{1-p_i} \right) \right] + \ln \left( \frac{T_{ii} }{T_{ij}} \right)}{\ln \left( \frac{p_i}{p_0} \left(\frac{1-p_0}{1-p_i}\right) \right)} \label{eqn:McondVitEvent}
\end{equation}
where $p_i$ is the probability of emitting a high-power Fourier bin given hidden state $i$. The resulting detection criterion is of the same linear form as Equation \ref{eqn:McondVit}, though with an altered final term, depending on the transition matrix elements of the proposed scatter.

In most cases, the direct event rate to a given frequency bin is significantly smaller than the rate of scattering to that bin, given an already present electron event, yielding transition matrix elements more sensitive to scattered tracks.
Detection in the vicinity of another long track therefore has a reduced threshold, yielding greater sensitivity to short tracks.
This defines the detection decision regions illustrated in Figure \ref{fig:vittyDecision}, in which the incorporation of event topology information expands the region of detectable tracks. Short duration or low signal-to-noise ratio (SNR) tracks are classified by the Viterbi algorithm as being due to noise-only.

Equation \ref{eqn:McondVitEvent} also encapsulates the impact of different gas compositions, and their associated energy loss spectra, on signal detection via the transition matrix elements $T_{ij}$ ($i,j \neq 0$). 
Induction of the above derivation implies that Equation \ref{eqn:McondVitEvent} can be applied to any subsequent track in the event, consequently describing the Viterbi track detection condition for arbitrary event topologies.

This formulation of the HMM state structure assumes that at most one electron is present within the detector bandwidth at a given instant, which is not necessarily true.
In Project 8 Phase II $\mathrm{T}_2$ data, the observed event rate of approximately 1 per hour \cite{thesisChristine}, with each event lasting several milliseconds, implied that the probability of multi-electron coincidences was negligible.
Extensions to the above state space for multiple electrons are possible, though these are unlikely to provide meaningful improvements in the detection efficiency given the additional computational burden for most applications of CRES.

\subsection{Application to Raw Spectrograms}
\label{sec:raw}

Finally, one can reconfigure the Viterbi algorithm to reconstruct CRES signals directly in raw, continuous data, as opposed to the previously considered binary quantized sparse spectrograms. This is done by replacing the emission matrix, which stores the probability that each state emits each possible symbol, with a function containing the desired probability distribution function for an on-the-fly computation of this probability. Outside a constant penalty from removing a lookup table, such an infinite extension does not affect the overall computational time complexity.

We consider the typical case in which the noise spectrum is additive white Gaussian, as a result of thermal noise.
Given that the initial phases of the signals are unknown a priori, the voltage Fourier magnitudes for the noise-only and signal-plus-noise hypotheses are given by Rayleigh and Rician distributions \cite{bookRadarDetection}, respectively.

\begin{equation}
    f_\textrm{Rice}(y \mid \nu, \sigma) = \frac{y}{\sigma^2} \, e^{-\frac{y^2 + \nu^2}{2 \sigma^2}} \, I_0 \left( \frac{y \nu}{\sigma^2} \right)
    \label{eqn:rice}
\end{equation}
where $I_0$ is the modified Bessel function of the first kind, $\sigma^2$ is the thermal noise variance, $y$ is the voltage Fourier magnitude, and $\nu$ is the expected voltage amplitude under $\mathcal{H}_1$. The Rayleigh distribution, $f_\textrm{Rice}(y\mid 0, \sigma)$, corresponds to the voltage amplitude distribution under the $\mathcal{H}_0$ hypothesis. 

Using these continuous distributions for the emission probabilities, we can likewise express the Viterbi maximal-likelihood detection criterion for CRES signal detection within raw spectrograms:
\begin{equation}
        \sum^{N}_{n} \ln I_0 \left( \frac{y_n \nu}{\sigma^2} \right) > N \left[ \ln\left( \frac{T_{00}}{T_{ii}} \right) + \frac{\nu^2}{2 \sigma^2} \right]  \\
        + \cases{
        \ln\left( \frac{T_{00} \cdot T_{ii}}{T_{0i} \cdot T_{i0}}\right)  \quad \, N_\textrm{tr}=1 \\
        \ln\left( \frac{T_{ii}}{T_{ij}}\right)   \quad \qquad N_\textrm{tr}>1
        }
        \label{eqn:condRawRice}
\end{equation}
 where $N$ is the number of time bins in the track candidate and $N_\textrm{tr}$ is the total number of tracks in the event.

The versatility of this algorithm, namely that the same algorithm can be reapplied on different compression levels of the data series, is appealing from an experimental standpoint. For instance, one could use the sparse spectrogram, 1-bit version of the algorithm on a first pass through the data, in which data throughput is the limiting factor, followed by a higher resolution search on the track candidates, using the same algorithm applied directly to the raw spectrogram data.

In general, such as for non-thermal noise spectra, the Viterbi detection criterion is a sequential likelihood ratio test given by:

\begin{equation}
        \ln \mathcal{L}(\mathcal{H}_i, \mathcal{H}_0) >  N \ln\left( \frac{T_{00}}{T_{ii}} \right) 
        + \cases{ \ln\left( \frac{T_{00} \cdot T_{ii}}{T_{0i} \cdot T_{i0}}\right)  \quad \, N_\textrm{tr}=1 \\
        \ln\left( \frac{T_{ii}}{T_{ij}}\right)  \quad \qquad N_\textrm{tr}>1
        }
        \label{eqn:condDecisionGeneral}
\end{equation}
Equations \cref{eqn:McondVit,eqn:McondVitEvent,eqn:condRawRice} are specific instances of the likelihood for Johnson-Nyquist noise. The likelihood sum is evaluated on the time series in sequence, resetting to zero after the inequality is satisfied.

\begin{figure}[t]
\centering
\includegraphics[width=0.95\textwidth]{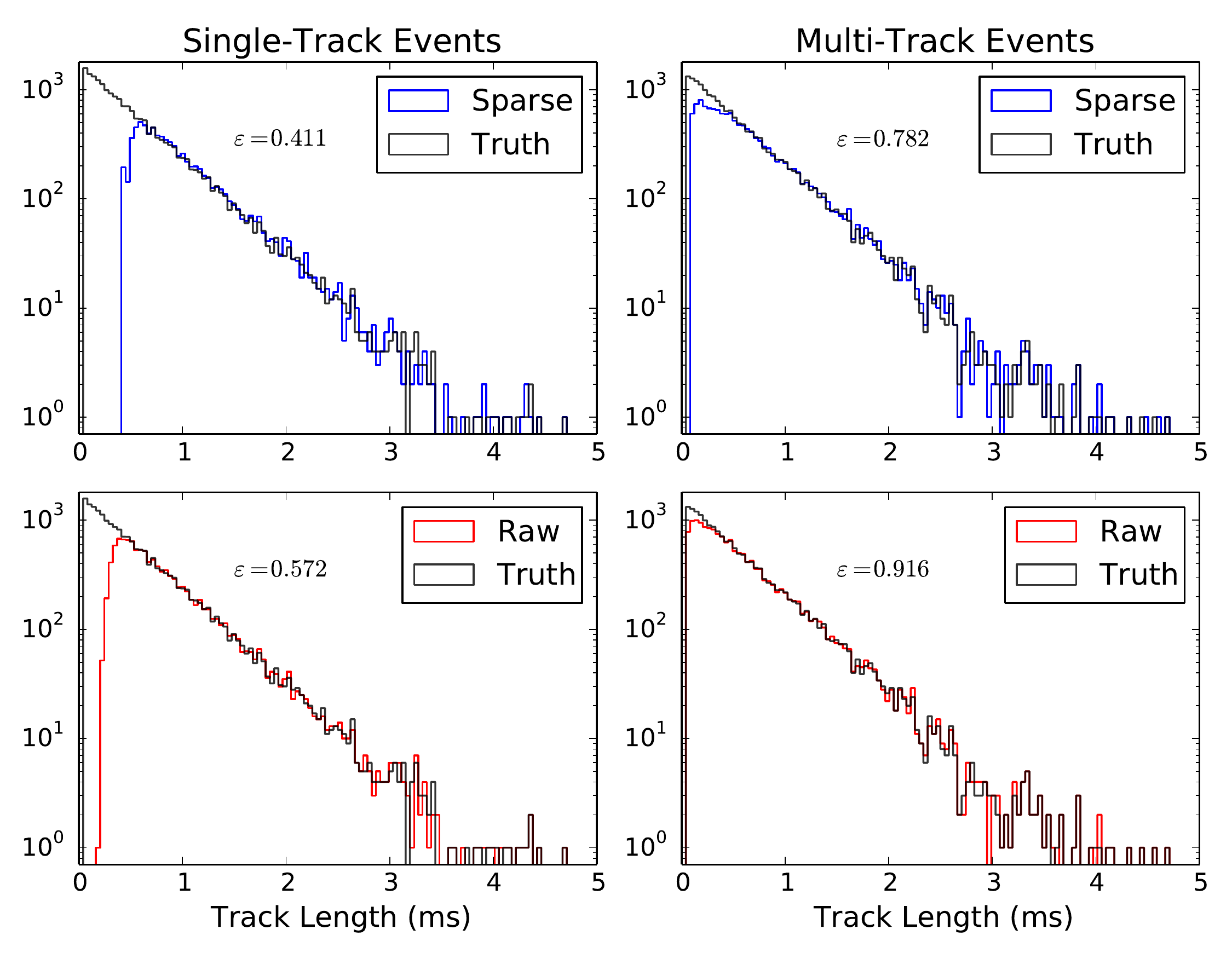}
    \caption{Track duration distributions of first reconstructed tracks in events from application of the Viterbi algorithm. Data generated using MC HMM data with Phase II detector parameters ($P_e=0.35$ fW, $T_N=135$ K, $\tau = 0.5$ ms), where $T_N$ is the system noise temperature. Illustrated and reconstructed for both binary quantized (sparse) and continuous (raw) Fourier data with $40.96 \, \mu \textrm{s}$ histogram time bins. MC reconstruction efficiency $\varepsilon$ inlaid for each event class.}
\label{fig:TLD_valid}
\end{figure}

\section{Results}

The application of the Viterbi algorithm to CRES signals is validated using both Monte Carlo (MC) studies of HMM generated data and Project 8 Phase II ${}^{83m}\mathrm{Kr}$ data \cite{artPhaseII,artKatrinEnergy}.
An example Viterbi track and event reconstruction is illustrated in Figure \ref{fig:vittySpectrograms}. In this 3-track Monte Carlo CRES event, the Viterbi algorithm correctly identifies and clusters the CRES tracks, despite the background noise. Without loss of generality, the inelastic scattering energy loss was uniformly distributed within three frequency bins, as opposed to the full effective energy loss spectrum resulting from the particular detector gas composition.

Figure \ref{fig:TLD_valid} illustrates the reconstruction performance of the Viterbi algorithm on simulated HMM MC data via the track duration distribution of first tracks in reconstructed events.
The underlying MC truth track duration distribution (black) is exponentially distributed given the time-independence of scattering in a constant density gaseous volume.
The resulting Viterbi reconstruction of MC HMM data is highly efficient for tracks longer than $\approx$1 ms, at the given signal and noise parameters.
Since temporally short tracks, with few high-amplitude bins, cannot be definitively distinguished from noise fluctuations, lower detection efficiency of short tracks is an inevitable consequence of false event rejection.

To decrease the computational burden of MC simulations, the event rate parameter $T_{01}$ is decoupled between event generation and reconstruction, such that the Viterbi decoding is applied to only event-dense data and not lengthy noise-only segments.
Similarly, in Project 8, finding the most probable hidden state sequence may not align exactly with the experimental goal of tritium endpoint neutrino mass inference. Even if a data segment is strictly more likely the result of a CRES electron than not (Equations \cref{eqn:condRawRice,eqn:condDecisionGeneral}), specific experimental goals may require greater rejection of false alarms.
If desired, a CRES experiment may use an unphysical $T_{01}$ in reconstruction for greater false alarm rejection.
Unphysical values for $T_{01}$ are equivalent to the general Neyman-Pearson decision rule $\ln \mathcal{L} > \eta$, which maximizes the detection probability for a given false alarm rate, which can be tuned with appropriate choice of $\eta$ ($T_{01}$).
Dependence of the false alarm rate on these Viterbi parameters lacks closed analytic form outside specific emission spectra \cite{bookQuickestDetection}.  

For the sparse spectrograms, the minimal detectable track length, 10 bins and 2 bins for the single-track and multi-track events respectively, is consistent with the Viterbi detection criteria at the simulated detector parameters. 
However, given the infinite domain of noise fluctuations, it is possible, though highly unlikely, that the raw spectrogram decision rule given by Equation \ref{eqn:condRawRice} can be satisfied within a single sample, so the ultimate limit on the quickest detection of CRES signals is ill-defined for raw data. Instead, Viterbi detection of CRES signals in noise yields a probability distribution function for the detection latency.
For non-negligible SNRs ($\nu^2/\sigma^2 \gtrsim 5$), Equation \ref{eqn:condRawRice} implies a median time required for detection:

\begin{equation}
    t_d = h \, \tau_\mathrm{{SNR}}
\end{equation}
where $h$ is the final term (braces) of Equations \cref{eqn:condRawRice,eqn:condDecisionGeneral}, and $\tau_\mathrm{{SNR}} = k_B T_N / P_e$, where $k_B$ is Boltzmann's constant, and $P_e$ is the radiated Larmor power. The shortest detectable track duration in this MC study was found to agree with the chosen threshold for Viterbi decoding, $t_d = 0.10$ ms. It should be noted that the median detection time is distinct from the mode of the track duration distribution, which scales with the mean free time of the electron in the gas. In addition, we note the displacement between the modes arising from differences in the reconstruction performance of single-track and multi-track events. No false events were found within the MC sample.

Direct application of the Viterbi algorithm to Phase II CRES data is challenging due to detector design choices which introduce additional signal complexity. 
In particular, electron-waveguide-mode interactions in Phase II can cause significant variations in the electron radiation rate, resulting in additional energy loss that varies on an event-by-event basis. To handle this, we apply a linear dechirp ($e^{-i \alpha t^2/2}$), $\alpha = 4\pi \cdot 10^{8} \space \textrm{s}^{-2}$,  removing most of the cyclotron-radiation-induced frequency slope from the spectrogram.
Additional energy losses of signals from cyclotron radiation are handled by the transition matrix in the same manner as scattering energy losses.
Significant variations in the event-by-event electron radiation rate are not expected to be a prominent feature of future phased-antenna array tritium endpoint neutrino mass CRES experiments, since these are limited to these small-volume waveguide detectors \cite{artP8IOP}.

\begin{figure}[t]
\centering
\includegraphics[width=0.95\textwidth]{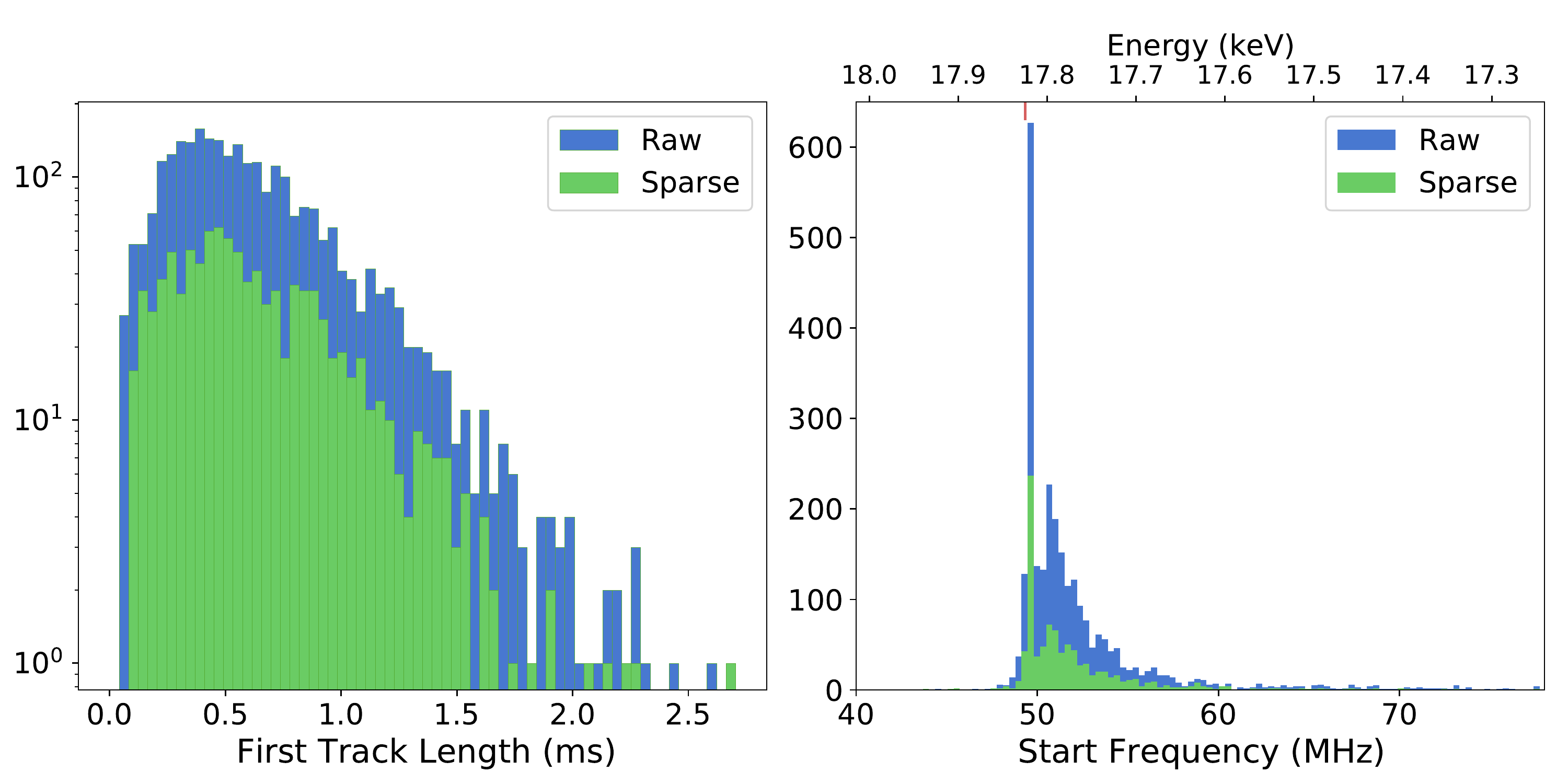}
    \caption{ Performance of Viterbi algorithm on Phase II ${}^{83m} \mathrm{Kr}$ sparse spectrogram data. Distribution of first track durations in reconstructed events with $40.96 \, \mu\textrm{s}$ histogram time bins (left), and start frequency distribution of reconstructed events (right). 17.825 keV ${}^{83m}\textrm{Kr}$ line highlighted in red.}
\label{fig:katy}
\end{figure}

Despite the simplified signal assumptions, such as fixed signal amplitudes, the Viterbi algorithm is applied to 10 hours of Phase II ${}^{83m}\textrm{Kr}$ data.
The Viterbi algorithm was implemented in C\texttt{++} using the Katydid \cite{gitKatydid} repository for standardized Project 8 file pre-processing.
Code optimizations, such as reducing the Viterbi search space by neglecting the possibility of electron energy gains from scattering reduced both time and memory costs by approximately 50\%.
Performance is assessed in Figure \ref{fig:katy} via diagnostics of reconstructed tracks and events.

Figure \ref{fig:katy} (left) illustrates the track duration distribution of the first track in events, which approximately mimics the idealized reconstruction performance for short tracks in MC simulations (Figure \ref{fig:TLD_valid}).
The right-hand plot of Figure \ref{fig:katy}, illustrating the initial frequency bin of the first track in reconstructed events, adds further support to the reconstruction performance, reproducing the detector response to the monoenergetic 17.825 keV K-32 line of ${}^{83m}\textrm{Kr}$.
In addition, since the Viterbi algorithm is applied to the frequency bandwidth 5--95 MHz, the absence of reconstructed events below $40$ MHz is consistent with the chosen $1/ 100$ day  false alarm rate. Use of the full information of the raw spectrogram yields an approximate $\times 3$ improvement in the detection efficiency compared to the sparse spectrogram Viterbi reconstruction, increasing the number of detected events from 976 to 2713 without additional apparent false alarms. 

Besides yielding analytical limits on the minimum track duration required for CRES signal detection, this work demonstrates that the Viterbi algorithm itself is a feasible and efficient reconstruction algorithm for CRES events.

\section{Conclusion}

CRES is a modern technique, well-suited for high-precision energy spectroscopy measurements of light charged particles, and of interest for next-generation electron spectroscopy experiments. Despite the potential experimental utility of the technique, a systematic examination of the general detection of CRES-like signals had not been performed directly. We used a novel application of the Viterbi algorithm, a dynamic programming algorithm which explicitly searches all possible hidden state sequences, in order to characterize maximal-likelihood CRES signal detection. 
Using the Viterbi decision rules, we can determine the expected minimal track duration that can be detected, as a function of the effective signal power and the system noise temperature, under arbitrary data quantization schemes and detector gas compositions.

\section*{Acknowledgments}

This material is based upon work supported by the following sources: the U.S. Department of Energy Office of Science, Office of Nuclear Physics, under Award No.~DE-SC0020433 to Case Western Reserve University (CWRU), under Award No.~DE-SC0011091 to the Massachusetts Institute of Technology (MIT), under the Early Career Research Program to Pacific Northwest National Laboratory (PNNL), a multiprogram national laboratory operated by Battelle for the U.S. Department of Energy under Contract No.~DE-AC05-76RL01830, under Early Career Award No.~DE-SC0019088 to Pennsylvania State University, under Award No.~DE-FG02-97ER41020 to the University of Washington, and under Award No.~DE-SC0012654 to Yale University; the National Science Foundation under Award Nos.~PHY-1806251 to MIT; This work has been supported by the Cluster of Excellence “Precision Physics, Fundamental Interactions, and Structure of Matter” (PRISMA+ EXC 2118/1) funded by the German Research Foundation (DFG) within the German Excellence Strategy (Project ID 39083149); Laboratory Directed Research and Development (LDRD) 18-ERD-028 at Lawrence Livermore National Laboratory (LLNL), prepared by LLNL under Contract DE-AC52-07NA27344, LLNL-JRNL-830010; the LDRD Program at PNNL; the University of Washington Royalty Research Foundation; and support from Yale University.  A portion of the research was performed using Research Computing at PNNL.  The isotope(s) used in this research were supplied by the United States Department of Energy Office of Science by the Isotope Program in the Office of Nuclear Physics.  We further acknowledge support from Yale University, the PRISMA+ Cluster of Excellence at the University of Mainz, and the Karlsruhe Institute of Technology (KIT) Center Elementary Particle and Astroparticle Physics (KCETA).		

\bibliographystyle{iopart-num}
\bibliography{biblio}

\providecommand{\newblock}{}
\begin{thebibliography}{10}
\expandafter\ifx\csname url\endcsname\relax
  \def\url#1{{\tt #1}}\fi
\expandafter\ifx\csname urlprefix\endcsname\relax\def\urlprefix{URL }\fi
\providecommand{\eprint}[2][]{\url{#2}}

\bibitem{artMonrealOG}
Monreal B and Formaggio J~A 2009 {\em Phys. Rev. D\/} {\bf 80}(5) 051301

\bibitem{artPRL0}
Asner D~M {\em et~al.\/} (Project 8 Collaboration) 2015 {\em Phys. Rev.
  Lett.\/} {\bf 114}(16) 162501

\bibitem{artKareem}
Kazkaz K and Woollett N 2021 {\em New J. Phys.\/} {\bf 23} 033043

\bibitem{artP8IOP}
Ashtari~Esfahani A {\em et~al.\/} 2017 {\em J. Phys. G Nucl. Part. Phys.\/}
  {\bf 44} 054004

\bibitem{bookRadarDetection}
DiFranco J~V and Rubin W~L 2004 {\em Radar Detection\/} (Scitech Publishing
  Inc.)

\bibitem{artP8Sensitivity}
Ashtari~Esfahani A {\em et~al.\/} 2021 {\em Phys. Rev. C\/} {\bf 103} 065501
  ISSN 2469-9993

\bibitem{artPhenoPaper}
Ashtari~Esfahani A {\em et~al.\/} 2019 {\em Phys. Rev. C\/} {\bf 99} 055501
  (\textit{Preprint} \eprint{1901.02844})

\bibitem{artViterbiOG}
Viterbi A 1967 {\em IEEE Trans. Inf. Theory\/} {\bf 13} 260--269

\bibitem{bookLightWavePapen}
Papen G~C and Blahut R~E 2019 {\em Lightwave Communications\/} (Cambridge
  University Press)

\bibitem{bookFSNLP}
Manning C and Sch{\"u}tze H 1999 {\em Foundations of Statistical Natural
  Language Processing\/} (MIT Press)

\bibitem{bookRNAViterbi}
Durbin R, Eddy S, Krogh A and Mitchison G 1998 {\em Biological Sequence
  Analysis: {Probabilistic} models of proteins and nuclei acids\/} (Cambridge
  University Press)

\bibitem{bookFundsConvCodes}
Johannesson R and Zigangirov K 2015 {\em Fundamentals of Convolutional
  Coding\/} 2nd ed (IEEE Press)

\bibitem{artHMMIntro}
{Rabiner} L and {Juang} B 1986 {\em IEEE ASSP Magazine\/} {\bf 3} 4--16 ISSN
  0740-7467

\bibitem{bookinfHMM}
Capp\'e O, Moulines E and Ryd\'en T 2005 {\em Inference in Hidden {Markov}
  Models\/} (Springer)

\bibitem{thesisChristine}
Claessens C 2020 {\em Event Detection in Project 8\/} Ph.D. thesis Johannes
  Gutenberg-Universit{\"a}t Mainz

\bibitem{Aseev2000}
Aseev V~N {\em et~al.\/} 2000 {\em Eur. Phys. J. D\/} {\bf 10} 39--52 ISSN
  1434-6079

\bibitem{artPhaseII}
Guigue M (Project 8 Collaboration) 2020 {\em J. Phys. Conf. Ser.\/} {\bf 1342}
  012025

\bibitem{artKatrinEnergy}
Altenmüller K {\em et~al.\/} 2020 {\em J. Phys. G. Nucl. Part. Phys.\/} {\bf
  47} 065002 ISSN 1361-6471

\bibitem{bookQuickestDetection}
Poor H~V and Hadjiliadis O 2009 {\em Quickest Detection\/} (Cambridge
  University Press)

\bibitem{gitKatydid}
Oblath N~S {\em et~al.\/} 2022 Katydid: Project 8 analysis software framework
  \urlprefix\url{https://github.com/project8/katydid}

\end{thebibliography}

\end{document}